\documentclass[journal]{IEEEtran}

\usepackage{amsmath,amssymb,amsfonts}
\usepackage{graphicx}
\usepackage{cite}
\usepackage{booktabs}
\usepackage{multirow}
\usepackage{array}
\usepackage{xcolor}
\usepackage{url}
\usepackage{hyperref}

\title{Self-Healing Coordination in Cognitive Swarm Agents with Bloch-Type Perceptual Memory}

\author{Jyotiranjan~Beuria%
\thanks{J. Beuria is with the IKS Research Center, ISS Delhi, India. E-mail: jyotiranjan.beuria@gmail.com.}%
}

\begin{document}

\maketitle

\begin{abstract}
Reactive flocking models usually map current local observations directly to motion, leaving limited room for internal perceptual state to shape recovery after disruption. Building on a non-Markovian collective-motion model based on self-regulated perceptual dynamics, we ask whether the Bloch-type slow-fast architecture can support self-healing coordination in cognitive swarm agents. Each agent carries a bounded Bloch-type perceptual register coupled to a slow regulatory state. The slow state is not treated as a standalone memory store; here, perceptual memory is used operationally to denote history-dependent cue resolution within the closed slow-fast loop. The Bloch update is a positivity-preserving effective dynamics for internal perceptual alternatives, not a microscopic quantum claim. We evaluate the architecture in a non-periodic, obstacle-rich drone migration task with finite speed, bounded turning, collision avoidance, altitude regulation, and a fixed migratory drive. Multi-seed ablations compare the full slow-fast architecture with memoryless and partial-feedback baselines using recovery time, largest-cluster restoration, polar order, local coherence, collision risk, and path efficiency. Results show that the main functional impact is on self-healing: after obstacle-induced fragmentation, the closed slow-fast loop accelerates restoration of spatial connectedness, whereas an uncoupled slow trace behaves like a memoryless controller.
\end{abstract}


\section{Introduction}

Collective coordination is a core problem for cognitive and developmental systems: each agent must convert partial local perception into group-level behavior without centralized supervision. Classical computational models have established that flock-like motion can emerge from simple local rules \cite{reynolds1987flocks}. Statistical-physics models show how alignment and noise can produce ordered motion and non-cohesive collective phases in self-driven particles \cite{vicsek1995novel,chate2008collective}. Biological studies refine this picture by showing that animal groups can exhibit spatial sorting and inferred interaction zones \cite{couzin2002collective,herbertread2011inferring}. Natural-flock measurements further connect collective motion to scale-free correlations, statistical-mechanical structure, and information transfer \cite{cavagna2010scale,bialek2012statistical,attanasi2014information}. Broader active-matter studies place these effects within a common language for self-propelled agents in complex environments \cite{marchetti2013hydrodynamics,bechinger2016active}.

These ideas have strongly shaped distributed multi-agent control. Nearest-neighbor models give conditions for heading agreement under changing interaction graphs \cite{jadbabaie2003coordination}, while flocking control adds cohesion and collision-avoidance structure to multi-agent motion \cite{olfati2006flocking}. Networked cooperation, graph-theoretic methods, and formation-control surveys clarify how topology and feedback structure affect group stability \cite{olfatisaber2007consensus,bullo2009distributed,mesbahi2010graph,oh2015formation}. In swarm robotics, the engineering focus is scalability and robustness: early surveys emphasize design methodology \cite{brambilla2013swarm}, formal treatments clarify collective mechanisms \cite{hamann2018swarm}, and recent reviews summarize applications and open challenges \cite{schranz2020swarm,dorigo2021swarm}. Physical robot swarms and collective construction systems show that large-scale behavior can be built from simple local programs \cite{rubenstein2014programmable,werfel2014designing}, while recent work on self-organizing nervous systems explores dynamic hierarchical organization for robot swarms \cite{zhu2024selforganizing}.

Drone swarms add constraints that are especially relevant for embodied cognitive agents: finite actuation, sensing limits, obstacle avoidance, and safety-critical motion. Agile micro-quadrotor studies demonstrate the feasibility of tightly coordinated aerial groups \cite{kushleyev2013towards}. Outdoor flocking experiments and optimized confined-environment drone swarms move these ideas toward real flight settings \cite{vasarhelyi2014outdoor,vasarhelyi2018optimized}. Aerial-swarm surveys and decentralized multirobot containment frameworks highlight the remaining difficulty of maintaining coherent group structure under local sensing and environmental constraints \cite{chung2018survey,hu2021decentralized}. The present paper focuses on the additional recovery question: after obstacles split a migrating swarm, can internal perceptual dynamics improve rejoining?

The important question is not only whether local rules can produce coordination, but whether internal agent state changes how coordination is recovered. Developmental or epigenetic robotics treats embodied behavior as a coupled process of perception, action, learning, and regulation \cite{asada2009survey,cangelosi2010developmental,cangelosi2015developmental}. Related work on bio-inspired embedded vision and occlusion-based coordination illustrates how perception mechanisms and task context can be integrated into autonomous robotic behavior \cite{hu2017bioinspired,hu2020occlusion}. This motivates a cognitive-agent view of swarm motion in which an agent is not merely a point particle with an instantaneous alignment rule. Instead, it is an embodied decision unit whose present action is shaped by recent perceptual reliability and self-regulatory state \cite{beuria2026nonmarkovian}.

The gap addressed here is that most flocking and swarm controllers remain effectively reactive: current neighbors and current obstacles determine the next action. Such controllers can align, avoid collisions, and form clusters, but they do not explicitly test whether internal perceptual dynamics can make recovery itself a closed-loop process. This distinction matters in open migration tasks, where obstacle rows can split a group into subgroups that remain highly aligned but spatially disconnected. We call the desired capability \emph{self-healing coordination}: the restoration of local coherence, polar order, and largest connected component after an externally induced split.

The model builds on a two-register non-Markovian framework in which each agent carries a fast perceptual register and a slow regulatory variable \cite{beuria2026nonmarkovian}. The fast register is implemented through bounded Bloch-type dynamics, while the slow variable integrates recent alignment and feeds back into future perceptual resolution. In this study, perceptual memory is used operationally to mean the history-dependent cue resolution of that closed perceptual-regulatory loop. The present paper does not attempt to re-establish the origin of memory itself. The proposed self-healing mechanism is precisely this slow-fast impact: Bloch-type perceptual resolution converts transient local cues into bounded alignment weights, and slow feedback carries their effect into subsequent steering during rejoining. The Bloch terminology is borrowed from driven-dissipative open-system models, where GKSL/Lindblad dynamics provide a standard positivity-preserving template \cite{gorini1976completely,lindblad1976generators,breuer2002theory,rivas2014quantum}. In this paper, however, the Bloch representation is an effective bounded-state model for internal perceptual alternatives, not a microscopic quantum claim.

We present a non-periodic, obstacle-rich drone migration task. The contributions of this work are: a cognitive-agent swarm controller coupling fast Bloch-type perceptual channels to slow regulatory feedback; an open-environment migration model with obstacles, cohesion, separation, altitude regulation, and bounded turning; self-healing metrics for fragmentation and recovery; and a multi-seed ablation that isolates the slow-fast loop from memoryless and partial-feedback baselines. Section~II defines the agent model, Section~III describes the migration task, Section~IV gives the metrics and protocol, Section~V reports the ablations, and Section~VI concludes.


\section{Cognitive-Agent Model}
\label{sec:model}

We consider a population of $N$ embodied agents moving in a two- or three-dimensional environment. Each agent is represented not only by its physical state, but also by internal perceptual and regulatory variables. The physical state of agent $i$ consists of position
\begin{equation}
    x_i(t)\in\mathbb{R}^{d},\qquad d\in\{2,3\},
\end{equation}
and a unit heading vector
\begin{equation}
    e_i(t)\in\mathbb{S}^{d-1},\qquad \|e_i(t)\|=1.
\end{equation}
The translational motion is fixed-speed:
\begin{equation}
    \dot{x}_i(t)=v_0 e_i(t),
    \label{eq:position}
\end{equation}
where $v_0>0$ is the nominal speed. The main modeling question is therefore how the heading $e_i$ is updated from local perception, emergent history dependence, and environmental steering cues.

\begin{figure*}[t]
    \centering
    \includegraphics[width=\textwidth]{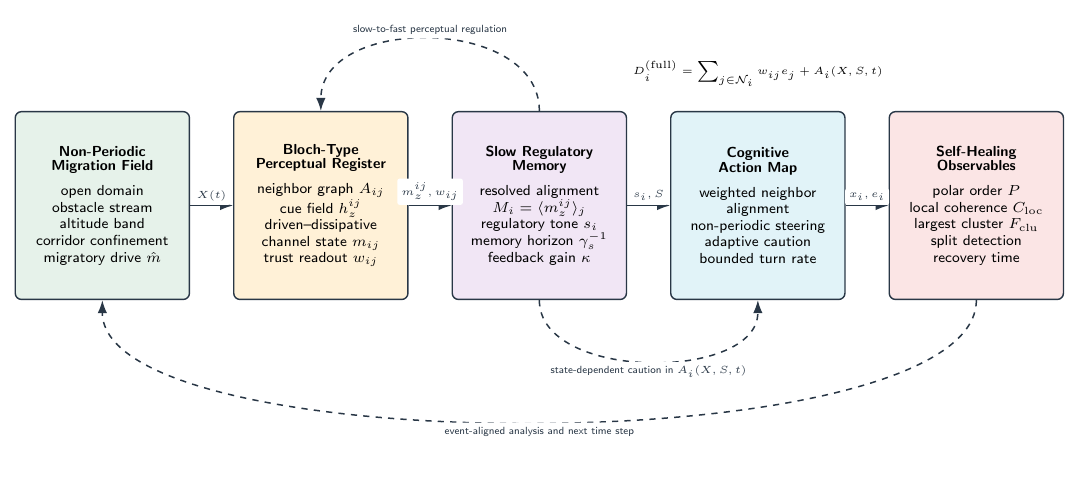}
    \vspace{-0.7em}
    \caption{Cognitive architecture of the proposed swarm-robotics controller. A non-periodic migration field supplies task and environmental context; local cues are resolved by Bloch-type perceptual registers; resolved alignment updates a slow regulatory state; regulatory feedback shapes later perceptual gain and adaptive steering; and self-healing observables quantify fragmentation and recovery.}
    \label{fig:cognitive_architecture}
\end{figure*}

\subsection{Neighbour Graph and Perceptual Channels}

At time $t$, agent $i$ interacts with agents inside a sensing radius $R$. The neighbour graph is defined by
\begin{equation}
    A_{ij}(t)=
    \begin{cases}
    1, & 0<\|x_i(t)-x_j(t)\|<R,\\
    0, & \text{otherwise}.
    \end{cases}
    \label{eq:adjacency}
\end{equation}
The neighbourhood of agent $i$ is denoted by
\begin{equation}
    \mathcal{N}_i(t)=\{j:A_{ij}(t)=1\}.
\end{equation}

Instead of assuming that agent $i$ instantaneously aligns with all neighbours, we assign a directed internal perceptual channel $(i,j)$ to each visible neighbour. This channel encodes the degree to which agent $i$ has resolved neighbour $j$ as an alignment-relevant cue. The internal fast register of channel $(i,j)$ is represented by a bounded Bloch-type vector
\begin{equation}
    m_{ij}(t)=
    \bigl(m^{ij}_x(t),m^{ij}_y(t),m^{ij}_z(t)\bigr),
    \label{eq:bloch_vector}
\end{equation}
where the longitudinal component $m^{ij}_z\in[-1,1]$ determines the resolved alignment tendency. The transverse components $m^{ij}_x$ and $m^{ij}_y$ represent unresolved switching and competition between internal alternatives. This representation is used as an effective bounded-state model for perceptual resolution, not as a claim of microscopic quantum dynamics.

\subsection{Fast Bloch-Type Perceptual Dynamics}

For each directed neighbour channel $(i,j)$, the perceptual register is driven by an effective field
\begin{equation}
    h_{ij}(t)=
    \bigl(\Gamma,0,h^{ij}_z(t)\bigr),
    \label{eq:field}
\end{equation}
where $\Gamma$ is the transverse switching rate. The longitudinal field is
\begin{equation}
    h^{ij}_z(t)
    =
    \kappa s_i(t)
    +
    g_{\mathrm{a}}\,
    e_i(t)\cdot e_j(t)
    +
    h^{ij}_{\mathrm{ext}}(t).
    \label{eq:hz}
\end{equation}
The first term is slow-to-fast regulatory feedback: the slow internal variable $s_i$ biases how strongly agent $i$ resolves local perceptual cues. The second term measures heading consistency between agents $i$ and $j$, with gain $g_{\mathrm{a}}>0$. The third term denotes possible external perceptual drive, such as goal direction, obstacle salience, or task-dependent cueing.

The fast register evolves according to a driven-dissipative Bloch-type update \cite{beuria2026nonmarkovian}:
\begin{align}
    \dot{m}^{ij}_x
    &=
    -2h^{ij}_z m^{ij}_y
    -
    \frac{m^{ij}_x}{T_2},
    \label{eq:mx}\\
    \dot{m}^{ij}_y
    &=
    2\left(h^{ij}_z m^{ij}_x-\Gamma m^{ij}_z\right)
    -
    \frac{m^{ij}_y}{T_2},
    \label{eq:my}\\
    \dot{m}^{ij}_z
    &=
    2\Gamma m^{ij}_y
    -
    \frac{m^{ij}_z-m^{ij}_{z,\mathrm{eq}}}{T_1}.
    \label{eq:mz}
\end{align}
Here $T_2$ is the transverse dephasing time and $T_1$ is the longitudinal relaxation time. The field-dependent equilibrium is chosen as
\begin{equation}
    m^{ij}_{z,\mathrm{eq}}
    =
    \tanh\left(h^{ij}_z\right),
    \label{eq:mz_eq}
\end{equation}
which keeps the resolved perceptual state bounded in $[-1,1]$. When $A_{ij}=0$, the corresponding channel is inactive and its contribution to alignment is set to zero.

The resolved alignment weight assigned by agent $i$ to neighbour $j$ is then
\begin{equation}
    w_{ij}(t)
    =
    \frac{1+m^{ij}_z(t)}{2},
    \qquad
    0\leq w_{ij}(t)\leq 1.
    \label{eq:weight}
\end{equation}
Thus, $m^{ij}_z=1$ corresponds to maximal alignment trust, $m^{ij}_z=0$ to neutral weighting, and $m^{ij}_z=-1$ to suppressed alignment.

\subsection{Slow Regulatory State}

Each agent also carries a slow regulatory variable
\begin{equation}
    s_i(t)\in\mathbb{R},
\end{equation}
which integrates recent resolved alignment. The mean resolved alignment perceived by agent $i$ is
\begin{equation}
    M_i(t)
    =
    \frac{1}{|\mathcal{N}_i(t)|}
    \sum_{j\in\mathcal{N}_i(t)}
    m^{ij}_z(t),
    \label{eq:mean_resolved_alignment}
\end{equation}
with $M_i=0$ when $\mathcal{N}_i$ is empty. The slow target is
\begin{equation}
    s^{\ast}_i(t)
    =
    \tanh\left(\lambda_{\mathrm{fb}} M_i(t)\right),
    \label{eq:s_target}
\end{equation}
where $\lambda_{\mathrm{fb}}$ is the fast-to-slow feedback gain. The regulatory variable relaxes according to
\begin{equation}
    \dot{s}_i
    =
    -\gamma_s\left(s_i-s_{\mathrm{base}}\right)
    +
    \gamma_s s^{\ast}_i.
    \label{eq:slow_memory}
\end{equation}
The regulatory history horizon is therefore controlled by $\gamma_s^{-1}$. Small $\gamma_s$ produces long-lived history dependence, while large $\gamma_s$ makes the internal state track recent alignment more quickly.

Equations \eqref{eq:hz} and \eqref{eq:slow_memory} form a closed slow-fast loop. The slow variable $s_i$ biases the fast perceptual field through $\kappa s_i$, while the fast longitudinal components $m^{ij}_z$ determine the slow target through $M_i$. This loop implements a minimal cognitive regulation mechanism: perception updates regulatory tone, and regulatory tone reshapes future perception. It is important to note that perceptual memory is the resulting history dependence of the closed loop, not an additional stored label. In a fragmentation event, the same loop provides a self-healing channel: bounded Bloch resolution prevents abrupt loss of cue continuity, while slow feedback sustains alignment sensitivity during post-obstacle rejoining.

\subsection{Heading Map and Bounded Actuation}

The resolved neighbour-alignment vector for agent $i$ is
\begin{equation}
    D^{\mathrm{align}}_i(t)
    =
    \sum_{j\in\mathcal{N}_i(t)}
    w_{ij}(t)e_j(t).
    \label{eq:alignment_vector}
\end{equation}
Environmental and task-dependent steering cues are collected in a vector
\begin{equation}
    A_i(X,S,t)\in\mathbb{R}^{d},
\end{equation}
where $X=\{x_i\}_{i=1}^{N}$ and $S=\{s_i\}_{i=1}^{N}$. This term may include separation, cohesion, obstacle avoidance, boundary regulation, altitude control, and migratory drive. The total desired direction is
\begin{equation}
    D_i(t)
    =
    D^{\mathrm{align}}_i(t)
    +
    A_i(X,S,t).
    \label{eq:desired_vector}
\end{equation}
The target heading is
\begin{equation}
    \hat{D}_i(t)
    =
    \frac{D_i(t)}{\|D_i(t)\|},
    \label{eq:target_heading}
\end{equation}
with $\hat{D}_i=e_i$ if $\|D_i\|$ is numerically zero.

To model finite actuation, the heading does not instantaneously jump to $\hat{D}_i$. Instead, it rotates toward $\hat{D}_i$ with bounded angular velocity:
\begin{equation}
    \angle\bigl(e_i(t+\Delta t),e_i(t)\bigr)
    \leq
    \omega_{\max}\Delta t.
    \label{eq:bounded_turn}
\end{equation}
Before this rotation, a small isotropic heading perturbation is added:
\begin{equation}
    \tilde{D}_i(t)
    =
    \frac{\hat{D}_i(t)+\eta \xi_i(t)}
    {\|\hat{D}_i(t)+\eta \xi_i(t)\|},
    \label{eq:noise}
\end{equation}
where $\xi_i$ is a random vector and $\eta$ controls perceptual or actuation noise. The heading update is therefore
\begin{equation}
    e_i(t+\Delta t)
    =
    \mathcal{R}_{\omega_{\max}\Delta t}
    \left(e_i(t),\tilde{D}_i(t)\right),
    \label{eq:heading_update}
\end{equation}
where $\mathcal{R}_{\theta}(e,\tilde{D})$ denotes rotation of $e$ toward $\tilde{D}$ by at most angle $\theta$ on the unit sphere.

\subsection{Limiting Cases}

The model contains standard collective-motion controllers as limiting cases. If the internal perceptual channels relax rapidly and the slow regulatory feedback is suppressed, then $m^{ij}_z$ approaches a static function of heading consistency,
\begin{equation}
    m^{ij}_z
    \approx
    \tanh\left(g_{\mathrm{a}}e_i\cdot e_j\right),
    \label{eq:fast_limit}
\end{equation}
and the heading update reduces to a weighted local-alignment rule. In the further weak-field limit,
\begin{equation}
    \tanh\left(g_{\mathrm{a}}e_i\cdot e_j\right)
    \approx
    g_{\mathrm{a}}e_i\cdot e_j,
\end{equation}
the controller behaves like a Vicsek/boids-type alignment mechanism with noise, cohesion, and separation.

If the transverse components are removed and only a scalar slow state is retained, the model becomes a slow-gain adaptive controller without internal perceptual competition. If both the slow variable and the fast register are removed, one obtains a purely reactive memoryless baseline:
\begin{equation}
    D_i^{\mathrm{reactive}}
    =
    \sum_{j\in\mathcal{N}_i} e_j
    +
    A_i(X,t).
    \label{eq:reactive_baseline}
\end{equation}
These limiting cases define the ablation models used in Section~IV. The full model differs from them by combining bounded perceptual resolution, transverse switching, dissipative relaxation, and slow regulatory feedback in a single closed-loop cognitive-agent architecture.

\subsection{Collective Observables}

The global polar order is
\begin{equation}
    P(t)
    =
    \left\|
    \frac{1}{N}\sum_{i=1}^{N}e_i(t)
    \right\|,
    \qquad 0\leq P(t)\leq 1.
    \label{eq:polar_order}
\end{equation}
The mean regulatory tone is
\begin{equation}
    S(t)
    =
    \frac{1}{N}\sum_{i=1}^{N}s_i(t).
    \label{eq:mean_tone}
\end{equation}
In obstacle-rich environments, global polar order alone can be misleading because two spatially separated subgroups may remain internally ordered while the swarm is fragmented. We therefore also use local coherence and largest-cluster fraction in later sections. These observables measure whether the group remains spatially connected and whether local neighbour headings are mutually consistent after environmental disruption.

\section{Non-Periodic Migration Environment}
\label{sec:environment}

The model in Section~\ref{sec:model} defines how Bloch-type perceptual regulation introduces history dependence into alignment. We now specify the non-periodic environment and steering field $A_i(X,S,t)$ used to test whether this internal regulation supports self-healing coordination after disruption. The task is designed to differ from periodic active-matter simulations in three ways: agents move in open space rather than on a torus, obstacles can split the group into spatially disconnected subgroups, and recovery must occur under finite actuation and local sensing.

\subsection{Steering Field Decomposition}

The environmental steering term in \eqref{eq:desired_vector} is decomposed as
\begin{equation}
    A_i(X,S,t)
    =
    A^{\mathrm{sep}}_i
    +
    A^{\mathrm{coh}}_i
    +
    A^{\mathrm{obs}}_i
    +
    A^{\mathrm{alt}}_i
    +
    A^{\mathrm{mig}}_i
    +
    A^{\mathrm{cor}}_i .
    \label{eq:steering_decomp}
\end{equation}
Here $A^{\mathrm{sep}}_i$ prevents short-range collisions, $A^{\mathrm{coh}}_i$ keeps the swarm spatially connected in open space, $A^{\mathrm{obs}}_i$ repels agents from obstacles, $A^{\mathrm{alt}}_i$ maintains a desired altitude band in three dimensions, $A^{\mathrm{mig}}_i$ imposes a task-level migratory direction, and $A^{\mathrm{cor}}_i$ confines the group to a broad migration corridor. Not all terms are active in every experiment. For example, altitude regulation is used only for $d=3$.

The steering field is modulated by the slow regulatory variable through an adaptive caution factor
\begin{equation}
    q_i(s_i)
    =
    1+
    c_{\mathrm{q}}\,
    \sigma(s_i),
    \qquad
    \sigma(s_i)=\frac{1}{1+\exp(-s_i)},
    \label{eq:caution_gate}
\end{equation}
where $c_{\mathrm{q}}\geq0$ controls how strongly regulatory tone increases avoidance gain. Thus, high regulatory tone makes an agent more cautious near obstacles, walls, and close neighbours.

\subsection{Short-Range Separation}

To avoid collisions, agents repel one another below a short distance $r_{\mathrm{sep}}<R$. Let
\begin{equation}
    r_{ij}=\|x_i-x_j\|,\qquad
    \hat{r}_{ij}=\frac{x_i-x_j}{\|x_i-x_j\|}.
\end{equation}
The separation force is
\begin{equation}
    A^{\mathrm{sep}}_i
    =
    q_i(s_i)
    \sum_{j\neq i}
    A_{ij}
    k_{\mathrm{sep}}
    \left(1-\frac{r_{ij}}{r_{\mathrm{sep}}}\right)_{+}
    \hat{r}_{ij},
    \label{eq:separation}
\end{equation}
where $(z)_{+}=\max(z,0)$ and $k_{\mathrm{sep}}>0$ is the separation gain. This term is active only when $r_{ij}<r_{\mathrm{sep}}$.

\subsection{Cohesion in Open Space}

In an open non-periodic domain, alignment alone does not guarantee spatial connectedness. We therefore include a weak local cohesion term that pulls each agent toward the centre of mass of its visible neighbours:
\begin{equation}
    \bar{x}_i
    =
    \frac{1}{|\mathcal{N}_i|}
    \sum_{j\in\mathcal{N}_i}x_j ,
    \label{eq:local_com}
\end{equation}
with $\bar{x}_i=x_i$ when $\mathcal{N}_i$ is empty. The cohesion term is
\begin{equation}
    A^{\mathrm{coh}}_i
    =
    k_{\mathrm{coh}}\left(\bar{x}_i-x_i\right),
    \label{eq:cohesion}
\end{equation}
where $k_{\mathrm{coh}}>0$ is chosen small enough that cohesion does not dominate the perceptual alignment dynamics. In the experiments, cohesion is treated as a necessary spatial constraint, while self-healing is evaluated by comparing the closed-loop perceptual architecture with memoryless models under the same cohesion gain.

\subsection{Obstacle Avoidance}

Obstacles are represented as circular or spherical exclusion regions. Obstacle $a$ has centre $o_a\in\mathbb{R}^{d}$ and radius $\rho_a$. Agents respond to an obstacle within a margin distance $\ell_{\mathrm{obs}}$ outside the obstacle boundary. Let
\begin{equation}
    d_{ia}^{\mathrm{obs}}
    =
    \|x_i-o_a\|-\rho_a,
    \qquad
    \hat{u}_{ia}
    =
    \frac{x_i-o_a}{\|x_i-o_a\|}.
\end{equation}
The obstacle-avoidance steering term is
\begin{equation}
    A^{\mathrm{obs}}_i
    =
    q_i(s_i)
    \sum_a
    k_{\mathrm{obs}}
    \left(1-\frac{d_{ia}^{\mathrm{obs}}}{\ell_{\mathrm{obs}}}\right)_{+}
    \hat{u}_{ia},
    \label{eq:obstacle}
\end{equation}
where $k_{\mathrm{obs}}>0$ is the obstacle gain. This term is strongest near the obstacle boundary and vanishes when the agent is farther than $\ell_{\mathrm{obs}}$ from the obstacle surface.

The important point is that obstacle avoidance is not treated as a separate centralized planner. It enters as another local steering cue in the same perception-regulation-action loop as alignment. When an obstacle divides the group, different subgroups receive different steering cues, producing temporary heading divergence and spatial fragmentation.

\subsection{Altitude Band for Drone-Like Agents}

For three-dimensional drone-like simulations, agents are constrained to remain inside a preferred altitude band
\begin{equation}
    z_{\min}\leq x_{i,z}\leq z_{\max}.
\end{equation}
The altitude correction is
\begin{equation}
    A^{\mathrm{alt}}_{i,z}
    =
    k_{\mathrm{alt}}
    \left[
    (z_{\min}-x_{i,z})_{+}
    -
    (x_{i,z}-z_{\max})_{+}
    \right],
    \label{eq:altitude}
\end{equation}
with horizontal components zero:
\begin{equation}
    A^{\mathrm{alt}}_{i,k}=0,\qquad k\neq z.
\end{equation}
This term prevents the swarm from escaping vertically while still allowing local altitude variation during obstacle avoidance.

\subsection{Migratory Drive}

The migration task imposes a fixed global direction
\begin{equation}
    \hat{m}\in\mathbb{S}^{d-1},
\end{equation}
analogous to a long-range goal direction or biological migratory heading. The corresponding steering term is
\begin{equation}
    A^{\mathrm{mig}}_i
    =
    k_{\mathrm{mig}}
    \left(1+n_i\right)\hat{m},
    \qquad
    n_i=|\mathcal{N}_i|,
    \label{eq:migration}
\end{equation}
where $k_{\mathrm{mig}}>0$ is the migration gain. The factor $(1+n_i)$ keeps the magnitude of the migratory drive comparable to the scale of the local alignment sum in \eqref{eq:alignment_vector}. In the main three-dimensional experiments, the migratory direction is chosen as
\begin{equation}
    \hat{m}=(0,1,0),
    \label{eq:migration_direction}
\end{equation}
so that the group travels in the positive $y$ direction.

\subsection{Migration Corridor}

To force repeated interaction with obstacles rather than trivial lateral escape, the migration task includes a broad soft corridor in the lateral coordinate $x$. The corridor half-width is denoted by $L_{\mathrm{cor}}$. The lateral correction is
\begin{equation}
    A^{\mathrm{cor}}_{i,x}
    =
    -k_{\mathrm{cen}}x_{i,x}
    -
    \operatorname{sgn}(x_{i,x})\,
    k_{\mathrm{wall}}
    \left(1+|x_{i,x}|-L_{\mathrm{cor}}\right)_{+}.
    \label{eq:corridor}
\end{equation}
The first term is a weak centering spring toward the corridor midline. The second term acts as a soft wall outside the corridor. The remaining components are zero:
\begin{equation}
    A^{\mathrm{cor}}_{i,k}=0,\qquad k\neq x.
\end{equation}
The corridor is not a closed box: agents remain free to migrate indefinitely in the $y$ direction.

\subsection{Streaming Obstacle Field}

The obstacle environment is generated as a sequence of rows placed ahead of the swarm centroid. Let
\begin{equation}
    c(t)=\frac{1}{N}\sum_{i=1}^{N}x_i(t)
    \label{eq:centroid}
\end{equation}
be the swarm centroid. Obstacles with $y$ position far behind the centroid are removed, while new obstacles are placed at random positions ahead of the centroid:
\begin{equation}
    y_a \in \bigl[c_y(t)+y_{\mathrm{near}},\,
                  c_y(t)+y_{\mathrm{far}}\bigr].
\end{equation}
The lateral coordinate of each obstacle is sampled inside the corridor, and its altitude coordinate is sampled inside the preferred altitude band. This produces a fixed but continually refreshed obstacle field in the moving frame of the swarm.

The obstacle rows are spaced so that the swarm can partially recover between successive disruptions. Thus, the task repeatedly alternates between two phases: an obstacle-avoidance phase, in which the group may split into subgroups, and a post-obstacle recovery phase, in which the group may rejoin and restore coherence.

\subsection{Self-Healing Coordination Task}

The self-healing task is defined by three requirements. First, the swarm must continue migrating in the direction $\hat{m}$ without centralized path planning. Second, it must avoid obstacles and short-range collisions using only local steering cues. Third, after an obstacle-induced split, it must recover spatial connectedness and heading coherence.

This task is intentionally stronger than measuring polar order in a periodic box. A group can have high polar order while being spatially fragmented. Conversely, a group can remain spatially close while having poor local heading coherence. We therefore treat self-healing as a joint recovery of three quantities:
\begin{equation}
    P(t) \rightarrow 1,
    \qquad
    C_{\mathrm{loc}}(t) \rightarrow 1,
    \qquad
    F_{\mathrm{clu}}(t) \rightarrow 1,
    \label{eq:self_healing_goal}
\end{equation}
where $P(t)$ is global polar order, $C_{\mathrm{loc}}(t)$ is local heading coherence, and $F_{\mathrm{clu}}(t)$ is the largest-cluster fraction. These latter two metrics are defined in Section~\ref{sec:experiments}.

The key experimental question is whether the two-register cognitive-agent model changes these recovery quantities relative to memoryless or partial-feedback baselines under the same environmental constraints. In this sense, obstacle-rich migration provides a functional test of emergent perceptual memory: the slow-fast Bloch loop is meaningful only if it improves self-healing beyond what can be obtained from ordinary alignment, cohesion, and repulsion alone.

\section{Experimental Protocol and Evaluation Metrics}
\label{sec:experiments}

This section defines the simulation protocol used to test how Bloch-type perceptual memory affects recovery from obstacle-induced fragmentation. The evaluation is designed around a single functional question: after environmental disruption splits the group, does the slow-fast Bloch architecture alter the restoration of collective coherence relative to simpler reactive controllers?

\subsection{Simulation Protocol}

Each experiment begins with $N$ agents initialized inside a finite spatial region around the migration corridor. Initial headings are random unit vectors. The swarm is first allowed to self-organize for a warm-up period without dense obstacle rows. After this transient, obstacles are streamed ahead of the swarm centroid as described in Section~\ref{sec:environment}. The camera frame and obstacle-generation window move with the swarm centroid, but the agents themselves move in open space.

The main three-dimensional migration experiments use the following default setting unless otherwise stated:
\begin{equation}
    d=3,\qquad
    \hat{m}=(0,1,0),
\end{equation}
with an altitude band
\begin{equation}
    z_{\min}=10,\qquad z_{\max}=18.
\end{equation}
The task duration is divided into discrete steps of size $\Delta t$. At each step, the neighbour graph is updated, fast perceptual channels are propagated, the slow regulatory variable is updated, the desired heading is constructed, and the position is advanced using \eqref{eq:position}. Obstacles behind the swarm are removed and new obstacles are inserted ahead of the centroid so that the swarm repeatedly encounters obstacle rows.

Table~\ref{tab:default_params} lists the default parameter values used in the reference simulations. The main full-trajectory run uses $N=200$ agents and $10\,000$ time steps. Obstacles are streamed in rows with $2$-$3$ obstacles per row and row spacing sampled uniformly from $[30,40]$ distance units, leaving recovery space between successive disruptions.

\begin{table}[t]
\centering
\caption{Default simulation parameters for the obstacle-rich migration task.}
\label{tab:default_params}
\begin{tabular}{lll}
\toprule
Parameter & Meaning & Default value \\
\midrule
$N$ & number of agents & $200$ \\
$d$ & spatial dimension & $3$ \\
$v_0$ & fixed speed & $2.0$ \\
$R$ & sensing radius & $9.0$ \\
$\Delta t$ & integration step & $0.05$ \\
$\omega_{\max}$ & max. turn rate & $0.40$ \\
$\eta$ & heading noise & $0.10$ \\
$T_1$ & longitudinal relaxation time & $0.6$ \\
$T_2$ & transverse dephasing time & $0.3$ \\
$\Gamma$ & transverse switching rate & $1.0$ \\
$\kappa$ & slow-to-fast gain & $1.2$ \\
$\gamma_s$ & slow regulatory rate & $0.03$ \\
$\lambda_{\mathrm{fb}}$ & fast-to-slow gain & $2.0$ \\
$k_{\mathrm{coh}}$ & cohesion gain & $0.50$ \\
$k_{\mathrm{sep}}$ & separation gain & $12.0$ \\
$r_{\mathrm{sep}}$ & separation radius & $1.8$ \\
$k_{\mathrm{mig}}$ & migration gain & $0.40$ \\
$L_{\mathrm{cor}}$ & corridor half-width & $14.0$ \\
$k_{\mathrm{cen}}$ & corridor-centering gain & $0.20$ \\
obstacle rows & spacing and row size & $[30,40]$, $2$-$3$ \\
\bottomrule
\end{tabular}
\end{table}

\subsection{Ablation Models}
\label{subsec:ablation}
All models are evaluated with the same obstacle, cohesion, separation, corridor, altitude, and migration terms. They differ only in the internal perceptual-regulatory mechanism used to construct the alignment weights. The main ablation study uses three models, chosen to isolate whether the slow state must feed back into future perceptual resolution.

\subsubsection{Reactive memoryless baseline}
The first baseline removes the slow regulatory loop and fixes the slow state at its baseline value:
\begin{equation}
    \gamma_s=0,\qquad \lambda_{\mathrm{fb}}=0,\qquad \kappa=0.
    \label{eq:baseline_reactive}
\end{equation}
The fast perceptual channels still relax under heading consistency, but no accumulated slow state can modulate perception or avoidance gain. This baseline captures memoryless local coordination with the same geometry and steering terms.

\subsubsection{Partial-feedback baseline}
The second baseline allows the slow variable to integrate recent resolved alignment, but prevents that slow trace from affecting future perceptual resolution:
\begin{equation}
    \kappa=0,
    \qquad
    A_i(X,S,t)\rightarrow A_i(X,t).
    \label{eq:baseline_partial}
\end{equation}
This model tests whether storing a slow internal state is sufficient, or whether the slow state must close the loop by shaping later perception and control.

\subsubsection{Full two-register model}
The full model uses \eqref{eq:mx}-\eqref{eq:slow_memory} and the heading map \eqref{eq:alignment_vector}-\eqref{eq:heading_update}. Its desired direction is
\begin{equation}
    D_i^{(\mathrm{full})}
    =
    \sum_{j\in\mathcal{N}_i}
    \frac{1+m^{ij}_z}{2}e_j
    +
    A_i(X,S,t).
    \label{eq:baseline_full}
\end{equation}
The comparison among \eqref{eq:baseline_reactive}-\eqref{eq:baseline_full} isolates the contribution of closed-loop slow regulatory feedback beyond ordinary local steering and beyond an uncoupled slow trace.

\subsection{Fragmentation and Recovery Events}

At each time step, we construct an undirected spatial graph using the same sensing rule as \eqref{eq:adjacency}. Let $\mathcal{C}_1(t),\ldots,\mathcal{C}_{K(t)}(t)$ be its connected components. The largest-cluster fraction is
\begin{equation}
    F_{\mathrm{clu}}(t)
    =
    \frac{1}{N}
    \max_{1\leq k\leq K(t)}
    |\mathcal{C}_k(t)|.
    \label{eq:largest_cluster_fraction}
\end{equation}
A fragmentation event is detected when
\begin{equation}
    F_{\mathrm{clu}}(t)<\theta_{\mathrm{split}}
    \label{eq:split_condition}
\end{equation}
after previously satisfying $F_{\mathrm{clu}}(t)\geq\theta_{\mathrm{split}}$. In the main experiments, $\theta_{\mathrm{split}}=0.85$, which treats only major fragmentation as a split event.

A recovery event is detected when the swarm returns to
\begin{equation}
    F_{\mathrm{clu}}(t)\geq\theta_{\mathrm{rec}},
    \qquad
    C_{\mathrm{loc}}(t)\geq\theta_{\mathrm{loc}},
    \label{eq:recovery_condition}
\end{equation}
and remains above these thresholds for a persistence window of $\tau_{\mathrm{hold}}$ simulation steps. Unless otherwise stated, $\theta_{\mathrm{rec}}=0.85$ and $\theta_{\mathrm{loc}}=0.85$. The recovery time for event $q$ is
\begin{equation}
    T_{\mathrm{rec}}^{(q)}
    =
    t_{\mathrm{rec}}^{(q)}
    -
    t_{\mathrm{split}}^{(q)}.
    \label{eq:recovery_time}
\end{equation}
If no recovery occurs before the next major obstacle encounter or before the end of the trial, the event is treated as unrecovered and reported separately.

\subsection{Coherence Metrics}

Global polar order is computed as in \eqref{eq:polar_order}. Since polar order alone can be high even for spatially separated groups moving in parallel, we also compute local heading coherence:
\begin{equation}
    C_{\mathrm{loc}}(t)
    =
    \frac{1}{|\mathcal{I}(t)|}
    \sum_{i\in\mathcal{I}(t)}
    \left\|
    \frac{1}{n_i(t)}
    \sum_{j\in\mathcal{N}_i(t)}
    e_j(t)
    \right\|,
    \label{eq:local_coherence}
\end{equation}
where
\begin{equation}
    n_i(t)=|\mathcal{N}_i(t)|,
    \qquad
    \mathcal{I}(t)=\{i:n_i(t)>0\}.
\end{equation}
If $\mathcal{I}(t)$ is empty, $C_{\mathrm{loc}}(t)$ is set to zero. This metric measures whether each agent's local neighbourhood has a consistent heading.

For each fragmentation event, we report the minimum coherence during the event,
\begin{equation}
    C_{\min}^{(q)}
    =
    \min_{t\in[t_{\mathrm{split}}^{(q)},t_{\mathrm{rec}}^{(q)}]}
    C_{\mathrm{loc}}(t),
    \label{eq:min_local_coherence}
\end{equation}
and the recovered coherence
\begin{equation}
    C_{\mathrm{post}}^{(q)}
    =
    \frac{1}{\tau_{\mathrm{post}}}
    \sum_{t=t_{\mathrm{rec}}^{(q)}}^{t_{\mathrm{rec}}^{(q)}+\tau_{\mathrm{post}}}
    C_{\mathrm{loc}}(t).
    \label{eq:post_recovery_coherence}
\end{equation}
Analogous quantities are computed for $P(t)$ and $F_{\mathrm{clu}}(t)$.

\section{Results and Ablation Analysis}
\label{sec:results}

This section evaluates the proposed model in two ways. First, a full $N=200$ trajectory with complete state recording is used to examine fragmentation, recovery, and internal-state dynamics in the revised rejoining regime. Second, a multi-seed ablation study compares the full closed-loop model against memoryless and partial-feedback baselines. We present some representative results below and additional details are available in a supplementary file.

\subsection{Representative Self-Healing Run}

\begin{table}[t]
\centering
\caption{Compact summary of the $N=200$ full closed-loop trajectory. Mean values refer to the average over frames.}
\label{tab:trajectory_summary}
\begin{tabular}{lc}
\toprule
Metric & Value \\
\midrule
Frames / agents & $10\,000$ / $200$ \\
Distance traveled & $944.4$ \\
$\overline{P}$ / $\min P$ & $0.956$ / $0.815$ \\
$\overline{C}_{\mathrm{loc}}$ / $\min C_{\mathrm{loc}}$ & $0.993$ / $0.956$ \\
$\overline{F}_{\mathrm{clu}}$ / $\min F_{\mathrm{clu}}$ & $0.711$ / $0.420$ \\
Mean regulatory tone $\overline{S}$ & $0.881$ \\
Split / recovered events & $5$ / $4$ \\
Mean recovered-event duration & $24.86$ \\
Mean active obstacles & $14.80$ \\
\bottomrule
\end{tabular}
\end{table}

\begin{figure}[t]
    \centering
    \includegraphics[width=\linewidth]{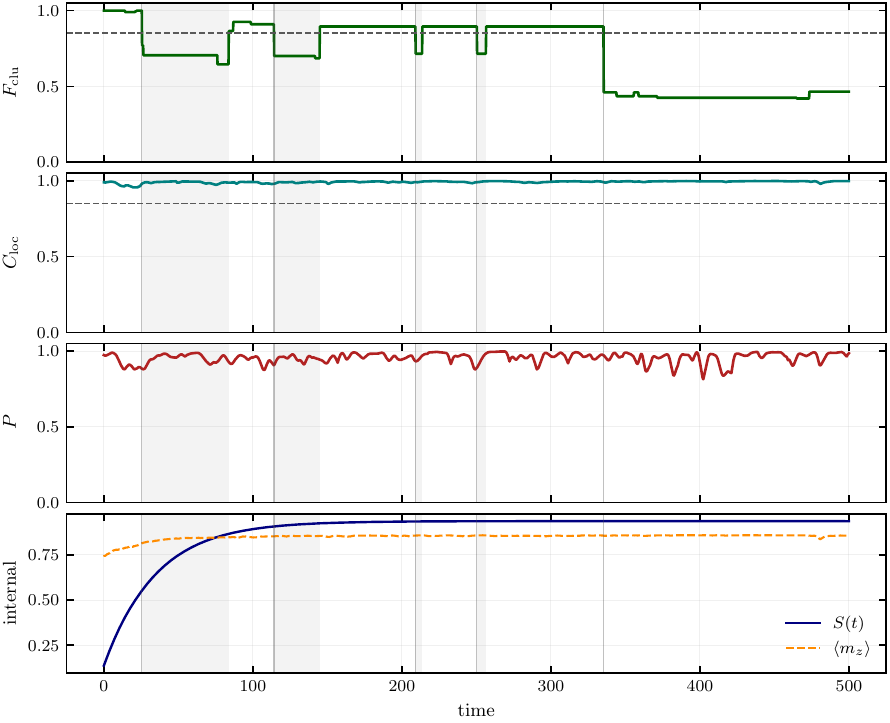}
    \vspace{-0.5em}
    \caption{Full-memory $N=200$ trajectory in the obstacle-rich migration task. Drops in $F_{\mathrm{clu}}$ identify spatial fragmentation; $C_{\mathrm{loc}}$ and $P$ remain high because separated subgroups often preserve local and global heading order.}
    \label{fig:main_dynamics}
\end{figure}

Figure~\ref{fig:main_dynamics} and Table \ref{tab:trajectory_summary} summarize the representative full closed-loop trajectory. The swarm traveled $944.4$ distance units over $10\,000$ frames with high mean polar order ($\overline{P}=0.956$) and high local coherence ($\overline{C}_{\mathrm{loc}}=0.993$). This confirms that the agents preserve directional coordination even in an obstacle-rich, non-periodic task. Spatial connectedness was more fragile: the mean largest-cluster fraction was $\overline{F}_{\mathrm{clu}}=0.711$, with a minimum of $0.420$. Thus, polar order alone would overstate recovery, because aligned subgroups can remain physically separated.

\begin{figure}[t]
    \centering
    \includegraphics[width=\linewidth]{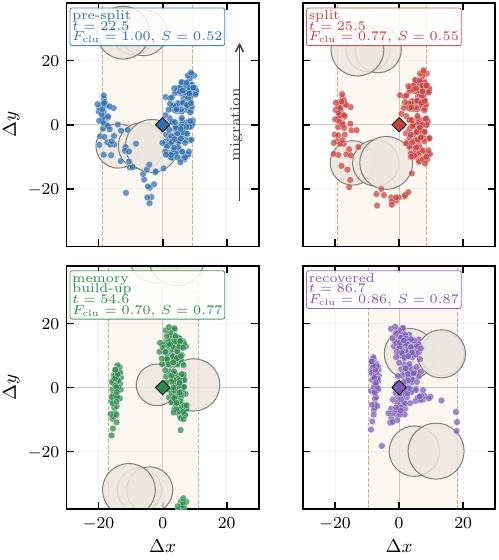}
    \vspace{-0.5em}
    \caption{Migration snapshots for the recovered event. Obstacle rows split the flock, after which local alignment, cohesion, corridor-centering, and closed-loop perceptual regulation gradually restore the largest connected component.}
    \label{fig:migration_snapshots}
\end{figure}

Five major split events were detected. Four recovered, giving a recovered-event fraction of $0.80$ and mean recovery time $24.86$. The first event (Figure \ref{fig:migration_snapshots}) began at $t=25.50$ and recovered at $t=83.70$, while later recovered events healed more rapidly. Across recovered events, the regulatory tone remained elevated during the split-recovery interval. This pattern suggests that the slow variable remains active through fragmentation and continues to shape post-obstacle coordination, but also shows that regulatory history is not a magic repair mechanism under severe spatial separation. 

Figure~\ref{fig:migration_snapshots} shows a representative recovered split event using four centroid-centered snapshots. In the pre-split frame, the swarm is nearly connected, with \(F_{\mathrm{clu}}=1.00\). As the obstacle row enters the local frame, obstacle avoidance divides the agents into two spatially separated but still coordinated subgroups, reducing \(F_{\mathrm{clu}}\) to \(0.77\). During the memory-build-up stage, the slow regulatory tone \(S\) continues to increase while the separated groups begin to reorient and compress toward the migration corridor. 


\subsection{Ablation Evidence and Mechanistic Interpretation}
We performed an ablation study to separate the effect of the closed slow-fast perceptual loop from ordinary alignment, cohesion, obstacle avoidance, and migration control. In particular, the memoryless and partial-feedback baselines (section \ref{subsec:ablation}) test whether recovery improves with functional feedback from the slow regulatory state, rather than merely with the presence of a slow internal trace or with shared reactive steering terms. We performed the simulation for 30 random seeds with $N=200$ and $T=3000$ steps for each of the three conditions. 


\begin{table}[t]
\centering
\caption{Ablation summary (mean$\pm$SEM) over $30$ random seeds and each run consists of 3000 frames.}
\label{tab:main_ablation}
\scriptsize
\setlength{\tabcolsep}{1.8pt}
\begin{tabular}{@{}lccccc@{}}
\toprule
Model & $\overline{F}_{\mathrm{clu}}$ & $\min F$ & $\overline{P}$ & Rec. & $T_{\mathrm{rec}}$ \\
\midrule
Full loop & $0.963\!\pm\!0.007$ & $0.852\!\pm\!0.026$ & $0.972\!\pm\!0.002$ & $18/23$ & $10.43\!\pm\!2.72$ \\
Memoryless & $0.957\!\pm\!0.009$ & $0.848\!\pm\!0.028$ & $0.972\!\pm\!0.002$ & $13/19$ & $14.76\!\pm\!3.67$ \\
Partial fb. & $0.957\!\pm\!0.009$ & $0.848\!\pm\!0.028$ & $0.972\!\pm\!0.002$ & $13/19$ & $14.76\!\pm\!3.67$ \\
\bottomrule
\end{tabular}
\end{table}

\begin{figure}[t]
    \centering
    \includegraphics[width=\linewidth]{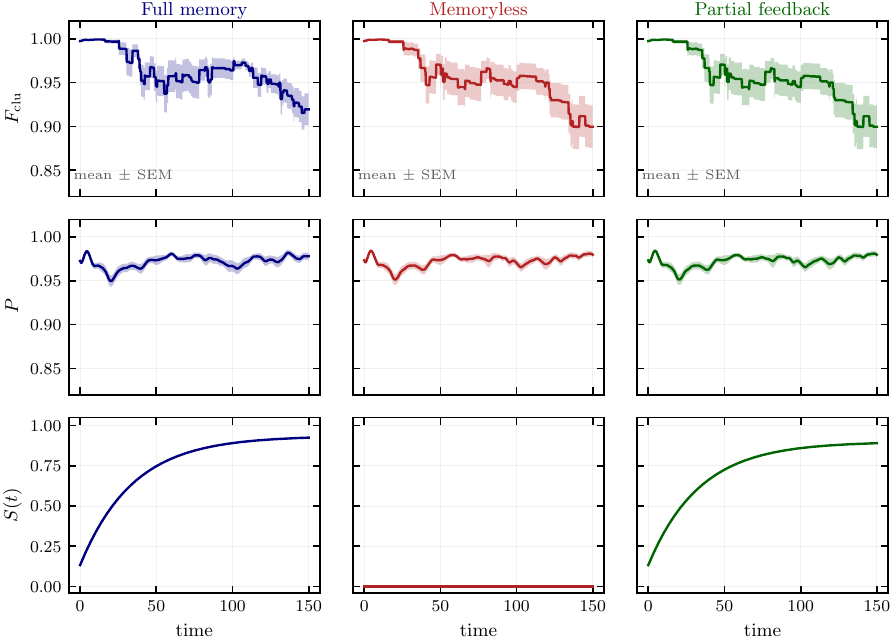}
    \vspace{-0.5em}
    \caption{Comparison of cluster fraction, polar order, and regulatory tone across ablations. The partial-feedback model stores a slow tone but lacks the slow-to-fast feedback needed to alter spatial robustness.}
    \label{fig:ablation_timeseries}
\end{figure}

\begin{figure}[t]
    \centering
    \includegraphics[width=\linewidth]{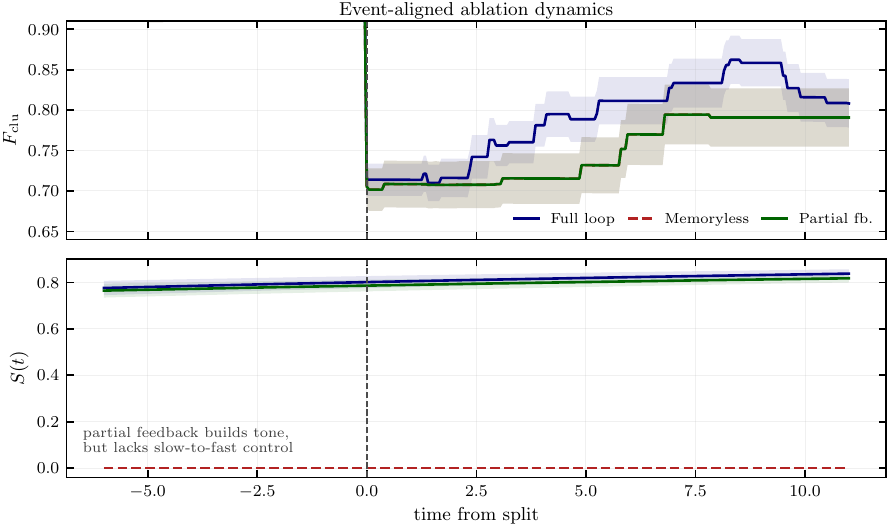}
    \vspace{-0.5em}
    \caption{Event-aligned recovery dynamics. Internal tone builds slowly across the split-recovery interval, while polar order can remain high even before full spatial reconnection.}
    \label{fig:mechanism}
\end{figure}

Table~\ref{tab:main_ablation} summarizes the self-healing effects averaged across the \(30\)-seed ablation study. The three controllers preserve comparable mean polar
order, indicating that all variants maintain directional alignment during migration.
The difference appears in the recovery-specific quantities: the full closed-loop
model recovers \(18/23\) detected split events, compared with \(13/19\) for both
the memoryless and partial-feedback baselines. It also reduces the mean recovery
time. Thus, the advantage of the full model is not a generic increase in flocking
order, but a more specific improvement in restoring spatial connectedness after
obstacle-induced fragmentation.

Figure~\ref{fig:ablation_timeseries} shows the same comparison at the time-series
level. The cluster-fraction traces reveal that spatial connectedness is the most
sensitive observable during obstacle-rich migration, whereas polar order remains
high and similar across models. The regulatory-tone panel provides the key
architectural control: the partial-feedback model can accumulate a slow internal
tone, but because this tone is not fed back into future perceptual resolution or
avoidance gain, its macroscopic behavior remains essentially memoryless. This
separates the presence of an internal state from its functional use in self-healing
coordination.

Figure~\ref{fig:mechanism} further aligns the trajectories around detected split
events. Immediately after a split, all models suffer a drop in
\(F_{\mathrm{clu}}\), but the full closed-loop model shows stronger post-split
restoration of the largest connected component. At the same time, the lower panel
shows that slow regulatory tone alone is not sufficient: partial feedback develops
\(S(t)\), yet its spatial recovery remains close to the memoryless case. Together,
Table~\ref{tab:main_ablation}, Fig.~\ref{fig:ablation_timeseries}, and
Fig.~\ref{fig:mechanism} support the central interpretation that self-healing
requires the closed slow-to-fast perceptual feedback loop, not merely stored
history or high heading alignment.




\section{Conclusion}
\label{sec:conclusion}

This paper developed a cognitive-agent model for obstacle-rich swarm migration in which each agent carries a fast Bloch-type perceptual register coupled to a slow regulatory state. The model was used as an effective, bounded, positivity-preserving internal-state dynamics rather than as a microscopic quantum description. By embedding this two-register mechanism into a non-periodic drone-like migration task, the study moved beyond periodic flocking and tested whether emergent perceptual memory can alter self-healing coordination after obstacle-induced fragmentation.

The results support a cautious but positive conclusion. In the representative full closed-loop trajectory, the swarm preserved high polar order and local coherence while undergoing repeated spatial fragmentation, recovering four of five major split events. In the multi-seed ablation study, the full closed-loop model improved recovered-event fraction and reduced mean recovery time relative to memoryless and partial-feedback baselines. The partial-feedback baseline stored a slow internal trace but behaved macroscopically like the memoryless controller, showing that perceptual memory becomes functional only when the slow trace feeds back into future perception and action.

The evidence does not imply that history dependence universally improves every observable. Average connectedness and polar order remain close across variants, and some closed-loop events fragment strongly before recovery. The main contribution is therefore more specific: Bloch-type perceptual dynamics acquire functional value when embedded in a slow-fast feedback loop, where they change the recovery pathway and can accelerate restoration of the largest connected component under non-periodic environmental disruption. This provides a bridge between active-matter models of collective motion and cognitive swarm robotics, where internal regulatory dynamics may be treated as testable self-healing mechanisms rather than passive histories.

\section*{AI Usage Declaration}
Generative AI tools are used to assist with language editing, formatting, and clarity during manuscript preparation. The scientific content, conceptual framework, analyses, and conclusions are developed by the author. The author takes full responsibility for the integrity, originality, and accuracy of the content of this manuscript.

\bibliographystyle{IEEEtran}
\bibliography{references}

\end{document}